\documentclass[aps,prb,preprint,groupedaddress,showpacs,showkeys,amsmath]{revtex4-1}
\usepackage[dvips]{graphicx}
\usepackage{dcolumn}
\usepackage{bm}
\usepackage{color}
\begin{document}

\title{Chemical accuracy from quantum Monte Carlo for the Benzene Dimer}

\author{Sam Azadi}
\email{s.azadi@ucl.ac.uk}
\affiliation{Department of Earth Science and Thomas Young Centre, 
University College London, London  WC1E 6BT, United Kingdom}
\author{R. E. Cohen}
\affiliation{London Centre for Nanotechnology, University College London, London  WC1E 6BT, and
Extreme Materials Initiative, Geophysical Laboratory, Carnegie Institution of Washington,
 Washington, DC, 20015, USA}
\date{\today}

\begin{abstract}
We report an accurate study of interactions between Benzene molecules using variational 
quantum Monte Carlo (VMC) and diffusion quantum Monte Carlo (DMC) methods.
We compare these results with density
 functional theory (DFT) using different van der Waals (vdW) functionals.
In our QMC calculations, we use accurate correlated trial wave functions
including three-body Jastrow factors, and backflow transformations. 
We consider two benzene molecules in the parallel displaced (PD) geometry, 
and find that by highly optimizing the wave function 
 and introducing more dynamical correlation into the wave function, 
we compute the weak chemical binding energy between aromatic rings accurately.
We find optimal VMC and DMC binding energies of -2.3(4) and   
-2.7(3) kcal/mol, respectively. The best estimate of the CCSD(T)/CBS limit is -2.65(2) 
kcal/mol [E. Miliordos {\it et al}, J. Phys. Chem. A {\bf 118}, 7568 (2014)]. Our results indicate 
that QMC methods give chemical accuracy for weakly bound van der Waals molecular interactions, 
comparable to results from the best quantum chemistry methods.
 
\end{abstract}
\maketitle

\section {Introduction}

Weak noncovalent van der Waals (vdW) interactions are fundamental to a wide range of topics 
relevance to physics, chemistry, and biology. A prototype vdW interaction 
is that resulting from the stacking between aromatic rings\cite{Beguin}.
 This interaction is crucially
important in biological systems such as protein folding\cite{Hunter1}, DNA's structure
and stability\cite{Cooper}. In addition,  aromatic rings  interactions play key roles in 
drug design\cite{Babine}, electronics\cite{Wang}, optical properties 
of materials\cite{Suponitsky},
polymer stability\cite{Pan}, conjugated carbon networks\cite{Sygula}, and crystal growth 
processes\cite{Portell,XWen}. 

In general vdW interactions are difficult to model accurately\cite{Burns,science14}.
Local and semilocal density functionals are unable to describe the long-range electronic 
correlation energy which is the main part of the vdW forces.
We use quantum Monte Carlo (QMC) and modern non-local exchange-correlation (XC) functionals.
Previously empirical and semi-empirical corrections were used. For instance,  
one approach was to add empirical, pairwise
atomic dispersion corrections of the form -C$_6/$R$^6$. To avoid double-counting
electron correlation effects at short range, these contributions can
be damped for small inter-atomic distances R. This method is  
referred to as DFT plus dispersion (DFT-D), and has been applied on different systems 
using various exchange-correlation (XC) funtionals  
\cite{Elstner,Wu,Tkatchenko1,Tkatchenko2,Ambrosetti,Reilly}. 

Less empirical approaches were also developed.
Effective nonlocal potentials were introduced\cite{Lilienfeld}, 
where the parameters were determined using fitting to \textit {ab initio} results. 
Becke and Johnson obtained dispersion
coefficients $C_6,C_8,C_{10}$ from the multipole moments\cite{Becke}. 
In their work, the moments were obtained from an electron and its exchange hole. 
The reliability of these approaches is similar to the DFT-D formalism. 

Another density functional based method, widely applied on noncovalent systems, is the combination 
of DFT with symmetry-adapted perturbation theory, refereed as DFT-SAPT or 
SAPT(DFT)\cite{Hesselman,Jeziorski,Parrish,TMParker}. In these approaches, 
the dispersion term is obtained using the frequency dependent density 
susceptibility function of time dependent DFT (TD-DFT). 
The perturbation theory of intermolecular interactions can accurately predict 
the complete intermolecular potential energy surfaces for weakly bound molecular
complexes\cite{Jeziorski}. From the point of view of perturbation theory, all the intermolecular
interactions, including van der Waals interactions, contain four fundamental
physical contributions: electrostatic, induction, dispersion, and exchange. 
The interactions differ only by proportions of these ingredients. The strongest 
of those interactions involve a larger negative contribution from the electrostatic 
forces as in hydrogen bonded systems. If the electrostatic contribution is small, 
like in interactions of rare gas atoms with molecules, the minima depths are 
often below 1 kcal/mol. Therefore, from this point of view, a system such as 
the benzene dimer is also a van der Waals complex.    

Finally, by including nonlocal terms in DFT correlation energy functional,
vdW-DFs\cite{vdW1,vdW2} include the long range 
nonlocal correlation energy obtained by the plasmon pole approximation.
These functionals were originally applied on different van der Waals systems to obtain 
potential energy curve (PEC)\cite{vdW2}. DFT-vdW functionals result in significant
improvements in equilibrium spacings between noncovalently
bound complexes, as well as in binding energy of weak interacting systems.  
The efficiency and 
accuracy of different DFT-vdW functionals on bulk systems were analyzed\cite{Graziano,Klimes1,Klimes2}.
The functionals are a clear improvement over semi-local functionals, although tests 
on a wider range of systems are desirable.

Numerous theoretical works have compared different approximate quantum chemistry   
based methods for noncovalent weakly bound systems\cite{SGrimme,Takatani,TTakatani,Avoird}.       
Particularly the coupled-cluster theory through perturbative triplets CCSD(T) which
is often considered as the gold standard for chemical accuracy\cite{Raghavachari,TJLee}. 
However, due to its substantial computational cost, scaling as $\text N^7$ 
where N is the number of electrons,
more efficient methods for vdW systems are highly desirable. 
These methods also cannot be applied to condensed matter. Thus we consider Quantum Monte Carlo (QMC),
and compare with non-local density functionals. 

Quantum Monte Carlo, which solves the electronic Schr\"{o}dinger equation
stochastically\cite{Matthew,PRL14,NJP,FS}, is an alternative approach to quantum  mechanical methods. 
Diffusion quantum Monte Carlo (DMC) provides 
accurate energies for vdW systems\cite{Cox,Ma,Yasmine,NABenedek,sorella,Korth,Dubecky}.
 DMC is also able to produce 
an accurate description of systems where many-body interactions play a key role\cite{Gillan,Dario}. 
In general QMC based methods are faster than the most accurate post-Hartree-Fock schemes for large number 
of particles N. The computational cost of QMC methods scales usually 
as $\text N^3$-$\text N^4$ depending on the method.

The benzene dimer has become a benchmark system for electronic structure methods
for systems where van der Waals interactions are important. Despite its simplicity 
the problem of identifying the global minimum structure is particularly challenging 
as there are only subtle differences in the binding energies of the different configurations.  
According to quantum chemistry results, two critical factors for the
 binding energy of the benzene dimer are basis set and electron correlation\cite{Miliordos}. 
Our previous comprehensive study of benzene molecules\cite{sam} 
illustrates the importance of basis set in QMC energy calculations.
Once the Jastrow factor is optimized by keeping fixed the Slater determinant,
we obtained a good description of the atomization energy of the benzene molecule 
only when the basis of atomic orbitals is large enough and close to the CBS limit.
In this work, we demonstrate that by using better trial wave functions and converged basis sets, 
we obtain a chemically accurate description of binding energy between aromatic rings.

\section {Computational Details}
We used the CASINO QMC code\cite{casino} with  a trial function of Slater-Jastrow (SJ) form,
\begin{equation}
\Psi_{T}({\bf R})=\exp[J({\bf R})]\det[\psi_{n}({\bf r}_i^{\uparrow})]\det[\psi_{n}({\bf r}_j^{\downarrow})],
\label{eq6}
\end{equation}
where ${\bf R}$ is a $3N$-dimensional vector that defines the positions
of all $N$ electrons, ${\bf r}_i^{\uparrow}$ is the position of the i'th
spin-up electron, ${\bf r}_j^{\downarrow}$ is the position of the j'th
spin-down electron, $\exp[J({\bf R})]$ is the Jastrow factor, and
$\det[\psi_{n}({\bf r}_i^{\uparrow})]$ and $\det[\psi_{n}({\bf
  r}_j^{\downarrow})]$ are Slater determinants of spin-up and spin-down
one-electron orbitals.  These orbitals were obtained from DFT
calculations using the plane-wave-based Quantum Espresso code\cite{QS}. 
We used the local density approximation (LDA) to generate the orbitals in the Slater determinant for the trial wave function.
We chose a very large basis-set cut-off of 200 Ry to guarantee convergence to the complete basis
set limit\cite{sam}. The plane-wave orbitals were transformed into a
blip polynomial basis\cite{blip,Parker}. The quality of the blip expansion
, meaning the fineness of the blip grid, can be improved by increasing 
the grid multiplicity parameter and consequently results in a greater 
number of blip coefficients. The value of this parameter in our work is 2.0. 
The local density approximation (LDA) 
pseudopotentials are generated using the {\it OPIUM} pseudopotential 
generation program\cite{opium}. 
We also checked that the Kleinman-Bylander\cite{Kleinmann} transformation did not 
generate ghost states.
In our DMC calculations, the pseudopotential energy was evaluated using a variational technique\cite{casula}.
We used DMC time steps of 0.01 a.u. and 0.04 a.u. and extrapolated the results linearly 
to zero time step.
 
The Jastrow factor is a positive, symmetric, explicit function of 
interparticle distances.
We used a Jastrow factor consisting of
polynomial one-body electron-nucleus (1B), two-body electron-electron
(2B), and isotropic three-body electron-electron-nucleus (3B) terms. 
The main approximation in fermionic QMC is the fixed node approximation.
To reduce this error, we used 
backflow transformation (BF) in our trial wave functions\cite{pablo}.
In the backflow transformation, the orbitals in the Slater determinant are
evaluated not at the actual electron positions, but a quasi-electron positions 
that are functions of all the particle coordinates. The backflow function, which 
describes the offset of the quasi-electron coordinates relative to the 
actual coordinates, contains free parameters to be determined by an 
optimization method. It allows the nodal surfaces to move within variational optimization, 
so with BF the QMC is no longer strictly fixed node.
However, the subsequent DMC computations use the nodal surface that were determined during the VMC step. 

In QMC calculations, correlated wave function can be obtained by 
replacing the single determinant by a sum 
over configuration state functions (CSFs), using BF transformations of the electronic 
coordinates or by using pairing wave functions \cite{MDBrown,NNemec,MCasula,JCP,Marchi}. 
Our BF transformation includes both electron-electron and electron-proton terms, given by
\begin{equation}
X_i(\{\mathbf{r_j}\})=\mathbf{r_i}+\xi^{(e-e)}_i(\{\mathbf{r_j}\})+\xi^{(e-P)}_i(\{\mathbf{r_j}\})
\end{equation}
where $X_i(\{\mathbf{r_j}\})$ is the coordinate of electron $i$  which depends on the configuration of the system
$\{\mathbf{r_j}\}$, $\xi^{(e-e)}_i(\{\mathbf{r_j}\})$ and $\xi^{(e-P)}_i(\{\mathbf{r_j}\})$ are electron-electron
and electron-proton backflow displacements of electron $i$, respectively, given by
\begin{equation}
\xi^{(e-e)}_i(\{\mathbf{r_j}\})=\sum_{j\neq i}^{N_{e}} \alpha_{ij}(r_{ij}) \mathbf{r}_{ij} 
\end{equation}
\begin{equation}
\xi^{(e-P)}_i(\{\mathbf{r_j}\})=\sum_{I}^{N_{P}} \beta_{iI}(r_{iI}) \mathbf{r}_{iI}
\end{equation}
where $\alpha_{ij}(r_{ij})$ and $\beta_{iI}(r_{iI})$ are polynomial functions of electron-electron 
and electron-proton distance, respectively, containing optimizable parameters.

We use two methods for wave function optimization:
variance minimization and energy minimization \cite{umrigar,toulouse}.  
The parameters of Jastrow and backflow are first optimized by variance
minimization at the variational Monte Carlo (VMC) level \cite{varmin1,varmin2,Gurtubay,Drummond,pterm}.
Since trial wave functions generally cannot exactly represent an eigenstate, the 
energy and variance minima do not coincide. Therefore energy minimization 
should produce lower VMC energies. 
We have found that the lower VMC  energies lead to lower DMC energies if we use backflow transformations.
This is due to improved many-body nodes as well as reduction in the errors 
induced by using non-local pseudopotentials.

We used the Quantum Espresso code\cite{QS} for DFT-vdW calculations
 with ultrasoft pseudopotentials\cite{rappe1990} and 
Perdew-Burke-Ernzerhof (PBE)\cite{PBE} exchange correlation functionals.
 The plane wave basis had a well-converged cut-off of 80 Ry.

\section {Results and discussion}

\begin{figure}
\centering
\includegraphics[width=0.5\textwidth]{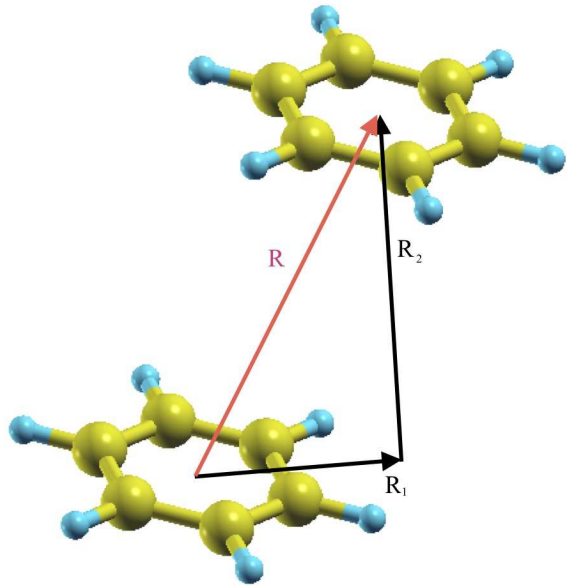}
\caption{\label{geom}(colour online) Parallel-displacement (PD) geometry of benzene dimer (C$_{2h}$ symmetry)
studied in this work. R indicates the distance between centers R = $\sqrt{( \text R_1^2 + \text R_2^2)}$. 
In our study centers displacement R$_1$ = 1.6 {\AA} is fixed.} 
\end{figure}

We used the experimental geometry for the benzene molecule, where 
the $C-C$ and $C-H$ bond lengths are 1.39 and 1.09 {\AA}, respectively.
Experiments support the existence of these three benzene dimer configurations\cite{Scherzer},
the parallel (sandwich), the T-shaped (C$_{2v}$), and the slipped-parallel or parallel-displacement
(PD, C$_{2h}$)\cite{Tsuzuki} configurations.
The T-shaped, where two benzene molecules are perpendicular to each other, and the PD configurations are more 
energetically favored than the parallel sandwich geometry 
\cite{Price,Hunter,Hobza1, Hobza2, Jaffe, Tsuzuki2, Tsuzuki3}.
T-shaped and PD configurations of benzene dimer are almost isoenergetic, and the benzene dimer potential 
energy surface is quite flat with several local minima separated by tiny barriers. 
In this work we focus on the PD configuration as shown in Figure ~\ref{geom}. 
Centers of two parallel benzene rings are displaced by R$_1$ = 1.6 {\AA}  and is fixed in our calculations. 
 
\begin{figure}
\centering
\includegraphics[width=1.0\textwidth]{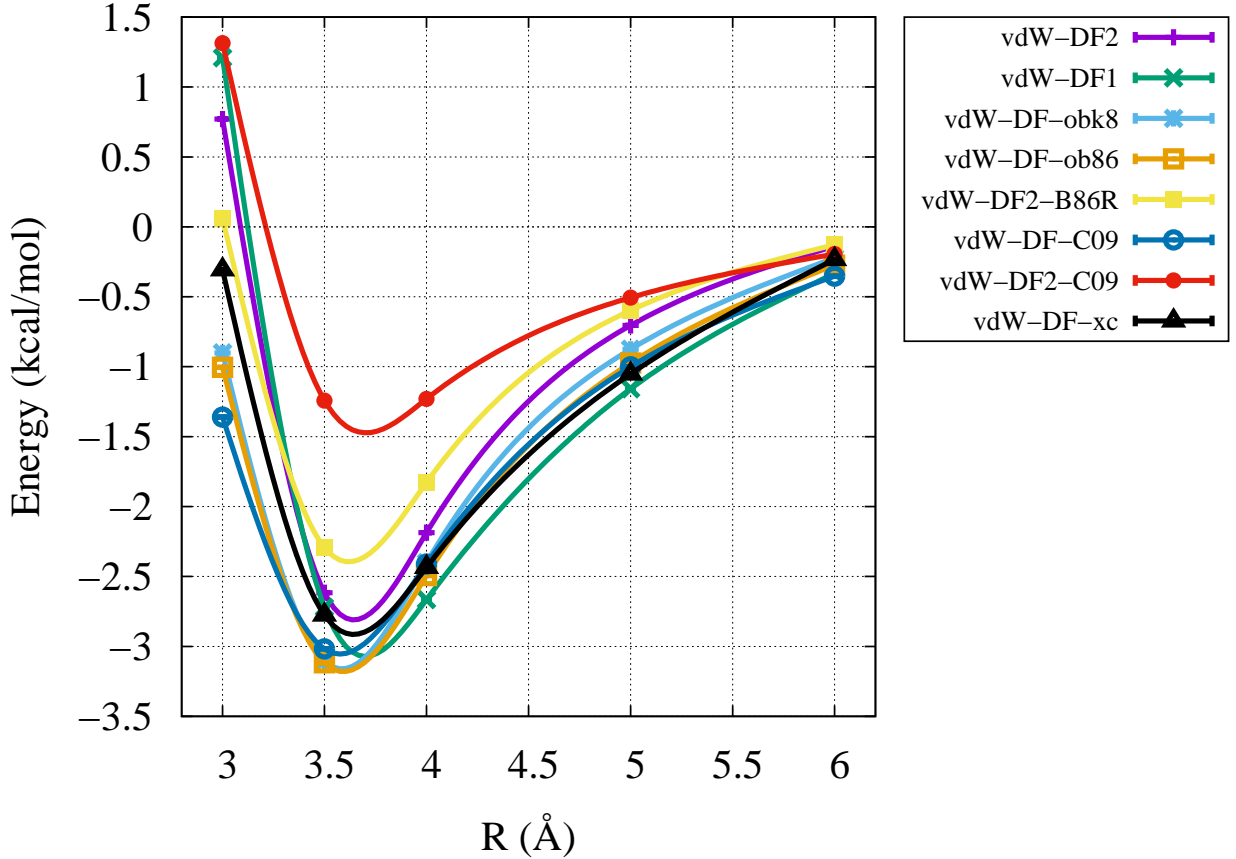}
\caption{\label{dft}(colour online) DFT energy of PD benzene dimer as a function of their separation 
obtained using different vdW-DF functionals. The reference is chosen at R = 10 {\AA}.
 Using polynomial fitting (Appendix), the optimal DFT binding 
energies obtained by vdW-DF1\cite{vdW1}, vdW-DF2\cite{vdW2}, vdW-DF-obk8, vdW-DF-ob86,
 vdW-DF2-B86R\cite{Klimes1,Klimes2}, vdW-DF-C09, vdW-DF2-C09\cite{C09},
and  vdW-DF-cx\cite{cx} are -3.1, -2.8, -3.1, -3.2, -2.4, -3.0, -1.5, -2.9 kcal/mol at R = 3.7, 3.65, 3.58, 
3.60, 3.63, 3.57, 3.71, and 3.65 {\AA}, respectively.
The experimental binding energy is -2.4(4) kcal/mol\cite{Grover}.}
\end{figure}

Figure ~\ref{dft} illustrates DFT potential energy curve obtained using different vdW functionals. 
We use vdW-DF1\cite{vdW1}, vdW-DF2\cite{vdW2}, vdW-DF-obk8, vdW-DF-ob86,
vdW-DF2-B86R\cite{Klimes1,Klimes2}, vdW-DF-C09, vdW-DF2-C09\cite{C09}, and vdW-DF-cx\cite{cx}
functionals.
 All vdW functionals use Slater exchange and PW\cite{PW} correlation functionals. 
The non-local terms are either vdW-DF1 or vdW-DF2. Employing various gradient correction on exchange energy
is the main difference between these functionals. 
 Using polynomial fitting (Appendix), the optimal DFT binding 
energies obtained by vdW-DF1, vdW-DF2, vdW-DF-obk8, vdW-DF-ob86, vdW-DF2-B86R, vdW-DF-C09, vdW-DF2-C09,
and  vdW-DF-cx are -3.1, -2.8, -3.1, -3.2, -2.4, -3.0, -1.5, -2.9 kcal/mol at R = 3.7, 3.65, 3.58, 
3.60, 3.63, 3.57, 3.71, and 3.65 {\AA}, respectively. 
 Free-energy landscape calculations using Car-Parrinello molecular meta-dynamics
methods using the BLYP density functional with dispersion corrections predict that T-shape geometry
is more stable at all temperatures\cite{Tummanapelli}. However, the PD configuration with C$_{2h}$ symmetry
has been determined using optical absorption spectroscopy, whereas a polar V-shape configuration with C$_{2v}$ 
symmetry has been suggested by multiphoton ionization mass spectroscopy\cite{Law, Bornsen}.      

\begin{table}[h]
\caption{\label{QMC}QMC energies (kcal/mol) of the PD benzene dimer obtained 
 by VMC and DMC using one-body (1B), two-body (2B),
 three-body (3B) Jastrow factors and Backflow(BF) transformations.
 Energies are calculated at different distance 
 geometry R {\AA}. Energy differences are calculated with respect to the large R = 10 {\AA}.
 Experimental binding energy is -2.4(4) kcal/mol \cite{Grover}. }
\begin{tabular}{ c c c c c c c }
\hline\hline
R   & VMC      & DMC      &   VMC   &  DMC      &  VMC    &    DMC  \\
    & 1B+2B    & 1B+2B    &  1B+2B  & 1B+2B     & 1B+2B   & 1B+2B  \\
    &          &          &  +3B    & +3B       & +3B+BF  & +3B+BF \\ \hline  
3.0 & 7.23(9)  & 5.63(9)  &  6.9(2) &  5.3(1)   &  5.0(2) &  4.4(3)   \\ 
3.5 & 2.50(9)  & -0.81(9) &  0.0(2) &  -0.9(1)  &  -1.2(2)&  -1.8(2)  \\ 
4.0 & 0.55(9)  & -1.45(9) &  -0.6(2)& -1.5(1)   &  -2.2(2)&  -2.5(2)   \\
5.0 & 0.02(9)  & -0.71(8) &  -0.0(2)& -0.8(1)   &  -0.6(2)&  -1.1(2)  \\
6.0 & 0.00(9)  & -0.31(9) &  0.0(2) &  -0.4(2)  &  -0.3(3)&  -0.4(2)  \\
\hline\hline
\end{tabular}
\end{table}

Table ~\ref{QMC} lists QMC energies of benzene dimer in the PD configuration obtained by 
VMC and DMC methods at different separation distance R. Lowest VMC and DMC energies 
are obtained at R = 4.0 {\AA}. Adding 3B-Jastrow factor substantially improve the VMC energies.
The 3B-Jastrow function takes care of what is missing in the 
1B and 2B Jastrow factors, meaning, the explicit dependence of the electron correlation 
on the ionic positions. At  R = 4.0 {\AA}, the difference between VMC energies obtained 
with and without 3B-Jastrow factor is about -1.1(2) kcal/mol, wheras in DMC
the difference is negligible. However, BF transformations 
significantly lower the QMC energies. At R = 4.0 {\AA}, 
BF transformations lower the energies by -1.6 and -1.0 kcal/mol       
at VMC and DMC calculations, respectively. Adding 3B-Jastrow factor and BF 
transformations improves the VMC and DMC energies by about -2.7(2) and -1.0(2) kcal/mol, 
respectively.

The VMC energies obtained with BF are lower than DMC energies without BF. This indicates 
that VMC with BF could be a useful level of theory to describe nonlocal long term 
interactions. In general, VMC calculations are significantly less expensive than 
DMC ones. Also, VMC has advantages for calculating expectation values of quantities 
more than the energy. Our QMC results clearly demonstrate that increasing the complexity 
of the wavefunction by including the BF correlations and 3B-Jastrow terms, plays key role 
in accurate describing of vdW interactions.  

\begin{figure}
\centering
\includegraphics[width=0.8\textwidth]{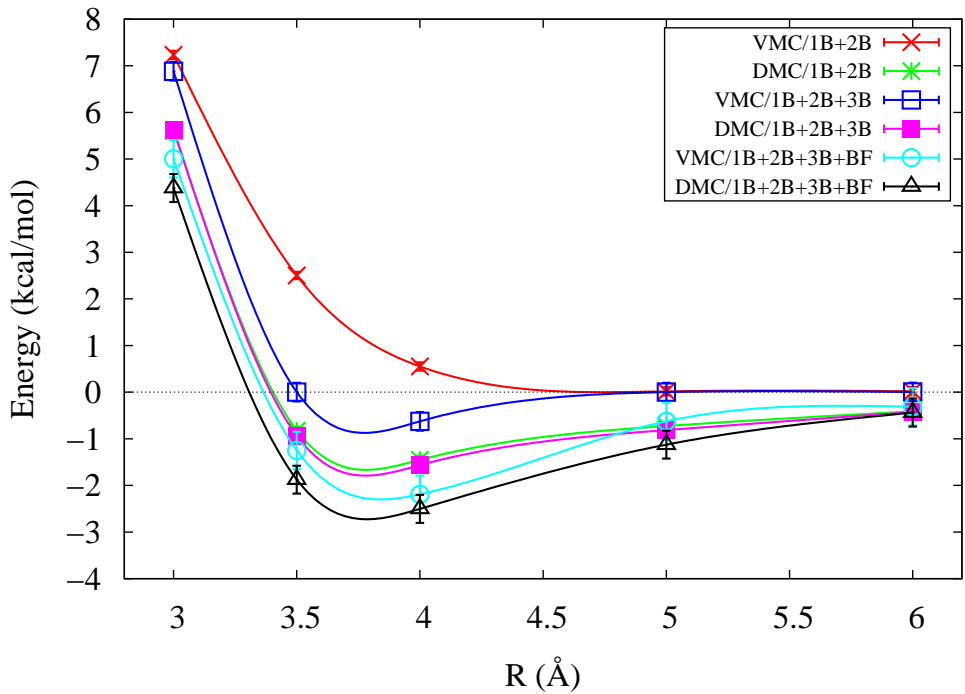}
\caption{\label{qmc_bind}(colour online) QMC energy of benzene dimer as a function of their distance 
obtained using VMC and DMC methods. Energy differences are calculated with respect to the large R = 10 {\AA}.
Using polynomial fitting (Appendix), the optimal values of binding energies obtained by DMC/1B+2B, VMC/1B+2B+3B,
DMC/1B+2B+3B, VMC/1B+2B+3B+BF, and DMC/1B+2B+3B+BF are -1.7(2), -0.9(2), -1.8(2), -2.3(4), and -2.7(3)
, respectively.}
\end{figure}

Figure ~\ref{qmc_bind} illustrates QMC potential energy curve of benzene dimer as a function of their 
distance obtained by VMC and DMC calculations. Using polynomial fitting (Appendix), 
the optimal values of binding 
energies at VMC level obtained by 1B+2B+3B-Jastrow and  1B+2B+3B-Jastrow plus BF 
correlations are -0.9(2), and -2.3(4) kcal/mol, respectively.
DMC optimal values obtained by 1B+2B-Jastrow, 1B+2B+3B-Jastrow, and  1B+2B+3B-Jastrow plus BF  
are  -1.7(2), -1.8(2), and -2.7(3) kcal/mol, respectively.
Using only 1B+2B-Jastrow factor, VMC is unable to provide a bound benzene dimer.
Using the same Jastrow factor the DMC binding energy 
is close to those ones obtained by SOS-MP2 method\cite{Miliordos}. 
Employing 3B-Jastrow factor significantly 
improves the VMC binding energy, whereas it doesn't lower the DMC binding energy considerably.
At the variational level, 
the inclusion of a 3B-Jastrow term provides additional dynamical correlation into the wavefunction and it is 
essentially useful for studying nonlocal vdW interactions. 

Although a 3B-Jastrow factor improves the binding energy of benzene dimer in VMC, substantial enhancement 
is obtained by employing BF correlations. The optimal values indicate that 
VMC and DMC energies are improved by -1.4 and -0.9 kcal/mol, respectively.
It suggests that BF is effective at improving the nodal surface of benzene dimer in PD geometry.
Considering that the LDA wave function often has too many nodal pockets, it is conceivable that 
BF coordinate transformations could modify the number of nodal pockets of a wave function.      
     
\begin{table}[h]
\caption{Binding energies $E_b$ of the PD benzene dimer obtained at different level of theories.
The zero point vibrational energy is not included.}
\label{comparison}
\begin{tabular}{ c c c c }
\hline\hline
Method$|$Basis set & R ({\AA}) & $E_b$(kcal/mol) & References \\ \hline
VMC/1B+2B+3B+BF    & 3.9(3)    & -2.3(4)         & This work \\
DMC/1B+2B+3B+BF    & 3.8(3)    & -2.7(3)         & This work \\
DFT-D/BLYP$|$TZVP  & 3.486     & -2.88           & Ref.~\onlinecite{Pitonak} \\
vdW-DF2            & 3.65      & -2.8           & This work \\
vdW-DF1            & 3.7       & -3.1           & This work \\
vdW-DF-obk8        & 3.58      & -3.1           & This work \\
vdW-DF-ob86        & 3.60      & -3.2           & This work \\
vdW-DF2-B86R       & 3.63      & -2.4           & This work\\
vdW-DF-C09         & 3.57      & -3.0           & This work\\
vdW-DF2-C09        & 3.71      & -1.5           & This work\\
vdW-DFT-cx         & 3.65      & -2.9           & This work\\ 
CCSD(T)$|$CBS      & 3.9       & -2.65(2)        & Ref.~\onlinecite{Miliordos} \\
CCSD(T)$|$CBS($\Delta$aDZ) &N/A& -2.73           & Ref.~\onlinecite{TTakatani}  \\ 
MP2$|$CBS          & 3.66      & -5.00(1)        & Ref.~\onlinecite{Miliordos} \\
JAGP-LRDMC         & 4.1(2)    & -2.2(3)         & Ref.~\onlinecite{sorella}  \\
FNDMC              & N/A       & -1.65(42)       & Ref.~\onlinecite{Korth}\\
Experiment         & N/A       & -2.4(4)         & Ref.~\onlinecite{Grover} \\
Experiment         & N/A       & -1.6(2)         & Ref.~\onlinecite{Krause} \\
\hline\hline
\end{tabular}
\end{table}

Table ~\ref{comparison} shows the binding energy of benzene dimer in PD geometry obtained by
different methods. Since the benzene dimer is a standard test for high-level quantum 
chemistry methods for proper characterization of vdW interactions
there are many more results; a comprehensive comparison 
between different high-level \textit{ab initio} approaches is reported recently\cite{Burns}.   
They provide databases for noncovalent interactions. 49 bimolecular complexes
in 345 geometry configurations are partitioned into subsets based on bonding motif.
Benzene dimer is in the dispersion-dominated subset.
Our DMC result is close to CCSD(T) (Table ~\ref{comparison}), 
which is considered the gold standard for chemical accuracy. The complete basis set (CBS) 
limit can now be estimated more precisely in the CCSD(T) framework. 
Our VMC results is comparable 
with those ones obtained by LRDMC method using Jastrow-AGP wave function\cite{sorella}. The  
correlated antisymmetrized geminal power (AGP)\cite{JCP} is the particle number conserving 
version of the Bardeen-Cooper-Schrieffer (BCS) wave function. A singlet valence bond between 
two electrons of opposite spin is determined by a geminal function. This framework successfully 
applied to identify the Kekul\'{e} and Dewar contributions to the chemical bond of the benzene 
molecule\cite{Marchi}.

The T-shape configuration was studied before using QMC methods\cite{Dubecky}.
Using 1B, 2B, and 3B Jastrow factors, they found that the binding energy of 
T-shape configuration obtained by FN-DMC is -2.88(16) kcal/mol. An accurate study 
of noncovalent systems illustrates the importance of Jastrow factor optimization 
in obtaining reliable results\cite{Korth}. They investigated in detail all 
technical parameters of QMC simulations. They also have found that the binding 
energies for T-shape and PD configurations are -3.77(39) and -1.65(42) kcal/mol, respectively.
The CCSD(T) estimates of T-shape binding energy is -2.74 kcal/mol \cite{Korth}.
By considering the reduction of the binding energy due to the ZPE 
($\Delta$ZPE=0.37 kcal/mol)\cite{sorella}, our DMC energy obtained by 3B-Jastrow and 
BF correlations is in excellent agreement with experiment\cite{Grover}.
Among DFT results obtained by different vdW functionals, 
vdW-DF2-B86R energy is close to our VMC/1B+2B+3B+BF result. vdW-DFT-cx and 
vdW-DF2 energies are close to our DMC/1B+2B+3B+BF energy.
 
We found that the dependence of the DMC energies on the quality of the trial wave function 
is significant. Whereas the Jastrow factors keep electrons 
away from each other and essentially improve the trial wave function, they do not change 
the nodal surfaces. It has been argued that BF transformation and 3B Jastrow correlation 
arises as the next-order improvements to the standard Slater-Jastrow wave function \cite{MHolzmann}.   
As our results show, the importance of BF correlations within DMC calculations is that they alter
the nodal surface and can therefore be used to reduce the FN error.
However, more complexity of BF-WF comparing to SJ-WF causes additional computational cost in QMC 
calculations. One of the most expensive operations in QMC calculations is the evaluation of the 
orbitals and their first two derivatives. The evaluation of the collective coordinates in 
BF-WF introduce significantly more computational cost.
 Moreover, whereas QMC calculations with SJ-WF require only the value,
gradient, and Laplacian of each orbital $\psi$, BF calculations also require cross derivatives 
such as $\partial^2 \psi / \partial x \partial y$. The most important complicating 
factor arising from BF transformations is that they make each orbital in the Slater determinants 
depend on the coordinates of every particle. 

\begin{figure}
\centering
\includegraphics[width=0.8\textwidth]{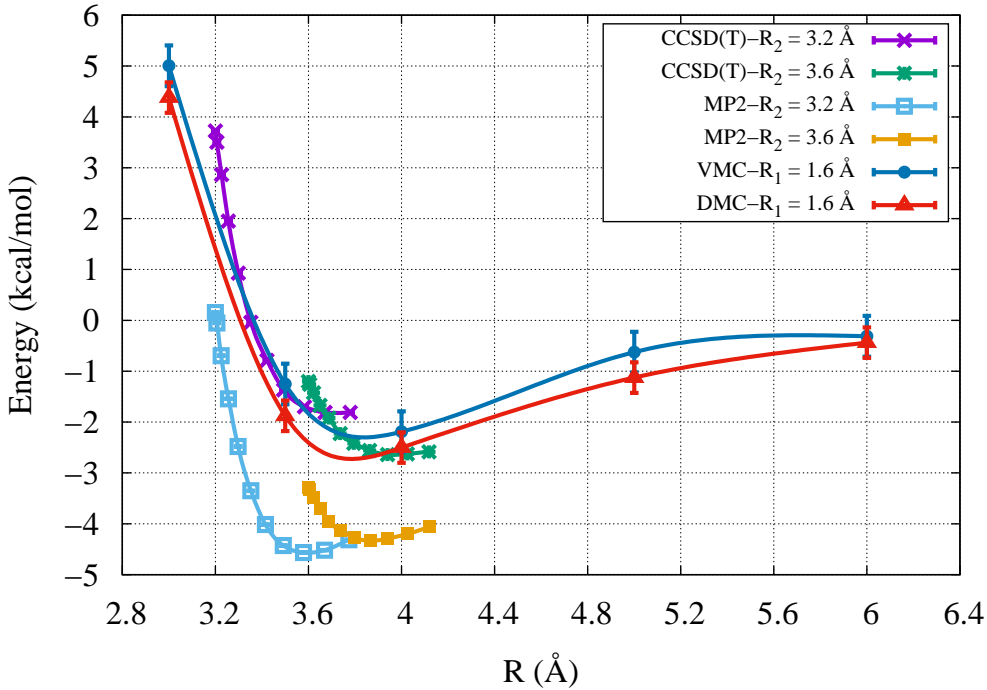}
\caption{\label{comp_bind}(colour online) Energy for PD benzene dimer as function of their 
centers distance obtained by different methods. We only compare our
VMC/1B+2B+3B+BF (VMC), DMC/1B+2B+3B+BF (DMC) results.
The MP2, and CCSD(T) data were taken from Ref ~\onlinecite{sinnok}.  } 
\end{figure}

Figure ~\ref{comp_bind} illustrates the potential energy curves for PD benzene 
dimer as function of ring centers distance at different level of theory. CCSD(T) and 
MP2 results were calculated using aug-cc-pVQZ$^{*}$
(=aug-cc-pVQZ - g function on carbon - f function on hydrogen)
 basis set\cite{sinnok}. At the
MP2 level, the equilibrium distances between aromatic rings are R$_1$ = 1.6 and R$_2$ = 3.4 
{\AA}. At the CCSD(T) level, they are R$_1$ = 1.6 and R$_2$ = 3.6 {\AA}. 
MP2 calculations produce over-binding. Comparing to 
T-shaped benzene dimer configuration, which we have not studied in this work, it 
has been claimed that PD configuration has larger electrostatic interactions\cite{sinnok},    
since the positive hydrogens on each aromatic ring are located
on top of the negative carbons of the other ring. In the PD geometry 
two benzene rings are closer together than in either the sandwich or the T-shaped 
configurations. Therefore, the incursion of the electronic orbitals of each ring 
make the electrostatic interaction more stable. 

\section{Conclusion}

We find that QMC can give chemical accuracy
for the benzene dimer. The good agreement among our results,
experiments and quantum chemistry methods, is an important sign of the capability
of the QMC based methods to provide accurate description of very weak intermolecular 
interactions based on vdW dispersive forces. We find that adding 3B-Jastrow terms and 
BF transformations leads to significant improvement in the accuracy of the weak 
vdW interaction between aromatic rings. BF-VMC energies are significantly lower than 
SJ-VMC and therefore BF-VMC could be useful alternative for a SJ-DMC calculations 
, which are more expensive. The accuracy of our VMC results compared 
with DMC is evidence of the high accuracy of our trial wave function. 
BF correlations give substantial enhancement in trial wave function of aromatic 
rings. By improving the nodal surface of wave function, it leads to a significant reduction 
in binding energy between two benzene molecules. Improved trial wave functions
will be useful in VMC calculations of quantities other than the energy, which are usually
more difficult to obtain accurately than the energy.    
We used single determinant wave functions in this work 
, but BF can also be combined with other types of wave functions such as 
multideterminant or pairing wave functions.  

\begin{acknowledgments}

This work made use of computing facilities provided by ARCHER,
the UK national super computing service, and by the University
College London high performance computing centre.
S. Azadi acknowledges that the results of this research have been achieved using the
PRACE-3IP project (FP7 RI-312763) resource ARCHER based in UK .
This work is supported by  the European Research Council (ERC) 
advanced grant ToMCaT (Theory of Mantle, Core and Technological Materials) and the Carnegie Institution of Washington.

\end{acknowledgments}

\newpage 

\section{Appendix: Energy curve fitting functions}\label{sec:app}

In this appendix we report fitting parameters.
We used piecewise polynomial fitting functions. 
Tables ~\ref{fit1} and ~\ref{fit2} list fitting parameters including
smoothing parameter $\alpha$, the sum of squares due to error (SSE), R-square, Root Mean Squared Error (RMSE),·¬
and optimal parameters (\textit{opt}) for each method.

\begin{table}[ht]
\caption{\label{fit1} Piecewise polynomial fitting parameters for QMC results. 
Smoothing parameter $\alpha$, the sum of squares due to error (SSE), R-square, Root Mean Squared Error (RMSE), 
and optimal parameters (\textit{opt}) for each method. Energies E and distance geometry R are in kcal/mol and {\AA}, respectively.}
\begin{tabular}{ c c c c c c c }
\hline\hline
         & VMC     & DMC     &   VMC   &  DMC    &  VMC     &  DMC     \\
         & 1B+2B   & 1B+2B   &  1B+2B  & 1B+2B   & 1B+2B    & 1B+2B    \\
         &         &         &  +3B    & +3B     & +3B+BF   & +3B+BF  \\ \hline  
$\alpha$ &0.9989716&0.9999155&0.9999155&0.9999155& 0.9998607&0.9998607  \\ 
SSE      &0.00372  &0.0002223&0.0002638&0.0002345& 0.0003988&0.0005139  \\ 
R-square &0.9999   &1.0000   &1.0000   &1.0000   & 1.0000   &1.0000     \\
RMSE     &0.1694   &0.1379   &0.1502   &0.1416   & 0.1443   &0.1638    \\
E$_{opt}$&0.00     &-1.66    &-0.87    &-1.77    & -2.31    &-2.71     \\
R$_{opt}$&4.50     &3.78     &3.78     &3.80     & 3.85     &3.79     \\
\hline\hline
\end{tabular}
\end{table}

\begin{table}[ht]
\caption{\label{fit2} Piecewise polynomial fitting parameters for DFT results. 
Smoothing parameter $\alpha$, the sum of squares due to error (SSE), R-square, Root Mean Squared Error (RMSE), 
and optimal parameters (\textit{opt}) for each method. Energies E and distance geometry R are in kcal/mol and {\AA}, respectively.}
\begin{tabular}{ c c c c c c c c c }
\hline\hline
         & vdW-    & vdW-    &   vdW-  &  vdW-   &  vdW-    &  vdW-   & vdW-     & vdW- \\
         & DF-     & DF-     &  DF2-   & DF-     & DF2-     & DF-     & DF1      & DF2  \\
         & obk8    & ob86    &  B86R   & C09     & C09      & cx      &          &  \\ \hline  
$\alpha$ &0.9996214&0.9996214&0.9996214&0.9996214& 0.9996214&0.9997703&0.99977035&0.99977035  \\ 
SSE      &0.001203 &0.001034 &0.0011   &0.0007068& 0.0008411&0.0003853&0.0007318 &0.0007439  \\ 
R-square &0.9998   &0.9998   &0.9997   &0.9998   & 0.9998   &0.9999   &0.9999    &0.9999   \\
RMSE     &0.1539   &0.1426   &0.1471   &0.1179   & 0.1286   &0.111    &0.1529    &0.1542  \\
E$_{opt}$&-3.13    &-3.16    &-2.37    &-3.03    & -1.46    &-2.90    &-3.06     &-2.79  \\
R$_{opt}$&3.58     &3.60     &3.63     &3.57     & 3.71     &3.65     &3.71      &3.65  \\
\hline\hline
\end{tabular}
\end{table}

\newpage

\nocite{*} 
\bibliographystyle{apsrev4-1}
\bibliography{JCP-rev-submit}

\end{document}